%% file: main.tex
\documentclass[prd, twocolumn, superscriptaddress,longbibliography,10.5pt]{revtex4-2}

\usepackage{xfrac}
\usepackage{amssymb}
\usepackage{comment}
\usepackage{caption}
\usepackage{xspace}
\usepackage{multirow}
\usepackage{ragged2e}
\usepackage{booktabs}
\usepackage{hyperref}
\usepackage{subcaption}
\usepackage{color}
\usepackage{mdwlist}
\usepackage[T1]{fontenc}
\usepackage{textcomp}
\usepackage{tgpagella}
\usepackage{amsmath}
\usepackage{hyperref}
\usepackage{lipsum}
\usepackage{sidecap}
\usepackage{graphicx}
\usepackage{axodraw4j}
\usepackage{bm}

\input{review_tool.tex}

\begin{document}

\title{High precision measurements of $\alpha_s$ at the future EIC}

\def\theaffn{\arabic{affn}} 
\input{authors_alphaS_rev}

\begin{abstract}
We present a projection study for the first moments of the inclusive spin structure function $\int g_1(x,Q^2) \, dx$ for the proton and neutron from simulated doubly-polarized $\vec{e}\vec{p}$ and $\vec{e}$-$^3 \overrightarrow{\rm He}$ collision data expected from the Electron Ion Collider (EIC). For detection and extraction of the neutron spin asymmetries from $\vec{e}$-$^3 \overrightarrow{\rm He}$ collisions, we used the double-tagging method which significantly reduces the uncertainty over the traditional inclusive method. Using the Bjorken sum rule, the projected results allow us to determine that the QCD coupling at the $Z$-pole $\alpha_s(M_{Z^0}^2)$ can be measured with a relative precision of $1.3\%$. This underscores the significance of the EIC for achieving precision determinations of $\alpha_s$.
\end{abstract}

\maketitle

\section{Introduction}

Quantum chromodynamics (QCD) is the quantum field theory (QFT) describing the strong force~\cite{Gross:2022hyw}, and one of the main components of the Standard Model (SM) of particle physics. The strength of QCD is characterized by its coupling constant: $\alpha_s$~\cite{Deur:2023dzc}, making it a key parameter of the SM.
Accurate knowledge of $\alpha_s$ is important for precisely calculating perturbative QCD (pQCD) series, as it serves as the expansion parameter. However, pQCD series convergence is slow due to the relatively large value of $\alpha_s$ and the renormalon problem~\cite{Deur:2016tte}. To match the precision required by present-day hadron scattering experiments that aim at testing the SM and exploring beyond the Standard Model physics, the relative uncertainty on $\alpha_s$ needs to be below a percent, lest it dominates the other uncertainties~\cite{Deur:2023dzc, dEnterria:2022hzv}.
Yet, achieving such experimental accuracy on $\alpha_s$ is a challenge. The world data compilation by the Particle Data Group results in $\Delta \alpha_s/\alpha_s = 0.85\%$~\cite{ParticleDataGroup:2020ssz}. This positions $\alpha_s$ as the least precisely known fundamental coupling when compared to the uncertainties on other fundamental couplings: $\Delta\alpha/\alpha=1.5\times 10^{-10}$ for the electromagnetic coupling, $\Delta G_F/G_F=5.1\times 10^{-7}$ for the weak coupling, and $\Delta G_N/G_N=2.2\times 10^{-5}$ for gravity's coupling~\cite{ParticleDataGroup:2020ssz}.
Consequently, a large effort is presently being devoted within the QCD community to reduce $\Delta \alpha_s/\alpha_s$~\cite{dEnterria:2022hzv}.

In this article, we study the accuracy at which $\alpha_s$ can be extracted from the Bjorken sum rule~\cite{Bjorken:1966jh, Adeva:1998} measured at the Electron Ion Collider (EIC)~\cite{Accardi:2012qut}.

We first discuss the extraction of $\alpha_s$ from deep inelastic scattering (DIS) observables, particularly focusing on the Bjorken Sum Rule (BJSR), which is formed from the nucleon spin structure functions $g_1^{p,n}$. Next, we present the simulation of EIC double-polarization data and projection studies on the double-spin asymmetry measurement. The projection is based on three energy settings each of $ep$ and $e$-$^3$He collisions with an integrated luminosity of 10~fb$^{-1}$ per setting. The polarization is assumed to be 70\% for both the electron and the proton (ion) beams. We then present projected results on the BJSR and extraction of 
$\alpha_s$ (evolved from the measured square 4-momentum transfer $Q^2$ to the $Z^0$ mass) 
and conclusions.

\section{Strong coupling from polarized DEEP INELASTIC SCATTERING: The Bjorken Sum Rule}

The underlying process of DIS, or inclusive inelastic lepton-hadron scattering at large $Q^2$ and large energy transfer, is elementary lepton-quark scattering.  DIS data are sensitive to $\alpha_s$ via violations of the Bjorken scaling, viz through the logarithmic $Q^2$-dependence of the nucleon structure functions~\cite{Bjorken:1969ja, Feynman:1969wa}.
At leading order (LO), scaling violations stem from gluon bremsstrahlung emitted by the struck quark of relative momentum $x$, photon-gluon fusion, and pair creations. At next-to-leading order (NLO), contributions arise also from the quark-photon vertex and quark self-energy.
Arguably, DIS structure functions are among the most robust observables for extracting $\alpha_s$ due to their fully inclusive nature, rendering them immune to uncertainties stemming from final-state hadronic corrections.

One avenue to obtain $\alpha_s$ from the $Q^2$-evolution of the structure functions is by concurrently fitting them (and possibly including other hard process data)~\cite{Deur:2023dzc,Deur:2016tte}. This is a complex endeavor involving global fits based on the Dokshitzer-Gribov-Lipatov-Altarelli-Parisi (DGLAP) evolution equations~\cite{Gribov:1972ri,Altarelli:1977zs,Dokshitzer:1977sg}. It also requires modeling nonperturbative inputs such as the quark and gluon distribution functions, and higher-twist (HT) corrections if data from low-$Q^2$ or high-$x$ regions are used as inputs. 

In the fitting method described earlier, data from double-polarization DIS experiments contributed minimally due to their scarcity at the necessary high-$Q^2$ values and their relatively large uncertainties compared to unpolarized data. Even with advancements like the EIC, polarized data will continue to have a limited impact on the extraction of $\alpha_s$ from global fits. This is because asymmetry measurements needed, e.g., for $g_1$ have necessarily lower statistics. Since the $Q^2$ (viz $\alpha_s$) dependence of polarized and unpolarized parton distribution functions (PDF) are similar,  these issues lessen the impact of polarized data on $\alpha_s$ extracted from global fits. Relative measurements (asymmetries) have the same issues and often employ PDFs as input to correct for unpolarized contributions, with a specific $\alpha_s$ value assumed to extract these PDFs thereby biasing its extraction. Furthermore, asymmetries have a suppressed $Q^2$-dependence. For these reasons, they are even less suited.  

On the other hand, double-polarization observables could contribute in a unique way through the $Q^2$-evolution of the first moments of spin structure functions~\cite{Deur:2018roz}, 
\begin{eqnarray}
\Gamma_1(Q^2) \equiv \int_0^{1^-} g_1(Q^2,x) \, dx~. \label{eq:Gamma1}
\end{eqnarray}
%
There is no model dependence aside from the generic assumption of SU(3)$_{\rm f}$ mass symmetry since the leading-twist (LT) nonperturbative inputs to $\Gamma_1$'s $Q^2$-evolution are the measured axial charges $a_0$, $a_3(\equiv g_A$) and $a_8$~\cite{Deur:2018roz}.
However, a hurdle is the unreachable low-$x$ part of the integral. The lowest $x$-value reachable depends on the beam energy, how forward (close to the beamline) the scattered leptons can be measured, and the minimum $Q^2$ value tolerable for data interpretation. The upcoming Electron Ion Collider will broaden the scope of double-polarization high-precision DIS measurements to encompass lower-$x$ regions previously unexplored.

%
In that context, it is beneficial to consider the isovector (proton minus neutron, p$-$n) combination $\Gamma_1^{\rm p-n}\equiv\int_0^{1^-} (g_1^{\rm p}-g_1^{\rm n}) \, dx$, called the Bjorken integral. 
%
It provides one of the two components of the 
BJSR~\cite{Bjorken:1966jh, Bjorken:1969mm} that links $\Gamma_1^{\rm p-n}$ to $g_A$ for $Q^2 \to \infty$ 
\begin{eqnarray} 
 \Gamma_1^{\rm p-n}(Q^2)\vert_{Q^2 \to \infty}{=} \frac{g_A}{6}.
\label{Eq:original bjorken SR}
\end{eqnarray} 
At finite $Q^2$ values, the pQCD processes mentioned above introduce an $\alpha_s$-dependence,
%
%
%

\begin{align} 
\Gamma_1^{\rm p-n}(\alpha_s ) = \Gamma_1^{\rm p-n}(Q^2 )  
=\sum_{\tau > 0} \frac{\mu^{\rm p-n}_{2\tau}(\alpha_s)}{Q^{2\tau-2}} \nonumber \\
=\frac{g_{\rm A}}{6}\bigg[1-\frac{\alpha_s(Q^2)}{\pi}
-3.58\left(\frac{\alpha_s(Q^2)}{\pi}\right)^2   \nonumber \\
-20.21\left(\frac{\alpha_s(Q^2)}{\pi}\right)^{3} 
 - 175.7\left(\frac{\alpha_s(Q^2)}{\pi}\right)^{4} - \nonumber \\
(\sim 893.38)\left(\frac{\alpha_s(Q^2)}{\pi}\right)^{5}
+  \mathcal O\left(\big(\alpha_{\rm {s}} \big)^6\right) \bigg]+\sum_{\tau>1} \frac{\mu^{\rm p-n}_{2\tau}(\alpha_s)}{Q^{2\tau-2}},
 \label{Eq:bjorken-SR}
\end{align} 
where the $\mu_{2\tau}$ are the coefficients of the HT expansion. Here, the series coefficients for the LT $\mu_2$ are calculated in the $\overline{\rm MS}$ renormalization scheme (RS)~\cite{Kataev:1994gd, Kataev:2005hv, Baikov:2008jh} and for $n_f=3$ quark flavors. Results on $\alpha_s$ extracted from Eq.~(\ref{Eq:bjorken-SR}) are therefore in $\overline{\rm MS}$ RS, which is the standard one for reporting $\alpha_s$~\cite{ParticleDataGroup:2020ssz}.
Since we aim at estimating the uncertainty $\Delta \alpha_s /\alpha_s$, $n_f$ may be kept fixed. For the actual extraction of $\alpha_s$ from data, quark
threshold effects should be accounted for both in the $\alpha_s$ series and the BJSR series, Eq.~(\ref{Eq:bjorken-SR}). Quark
threshold corrections for the latter are presently known at next-to-next-to
leading order (NNLO)~\cite{Blumlein:2016xcy}. We assume that they will be available for higher order by the time EIC delivers the Bjorken sum data. Finally and importantly, the determination of $\alpha_s$ using Eq.~(\ref{Eq:bjorken-SR}) does not assume $\alpha_s$, directly or indirectly in contrast to other approaches that, e.g., requires PDF inputs such as $\Delta G$ which are determined assuming a specific value of $\alpha_s$.

$\Gamma_1^{\text{p-n}}$ displays a relatively simple LT $Q^2$-evolution, known to a higher order than for the individual nucleon cases. This is crucial since the truncation of a pQCD series typically creates one of the dominant uncertainties when extracting $\alpha_s$~\cite{dEnterria:2022hzv}.
Equation~(\ref{Eq:bjorken-SR}) shows that the pQCD approximant of the BJSR is known at N$^4$LO (order $\alpha_s^4$), with an N$^5$LO estimate. The same accuracy is available for $\alpha_s$, for which the pQCD approximation has been calculated at five loops in the $\overline {\text{MS}}$ RS~\cite{Kniehl:2006bg}, i.e., up to $\beta_4$ in the QCD $\beta$-series~\cite{Deur:2023dzc}.
Additionally, the LT nonperturbative input is precisely measured, $g_A=1.2762(5)$~\cite{ParticleDataGroup:2020ssz}, and HT are known to be small for $\Gamma_1^{\text{p-n}}$~\cite{Deur:2014vea}. 
In Eq.~(\ref{Eq:bjorken-SR}) we only wrote the $Q^2$-dependence of the LT, $\mu_{2}$. 
The twist-4 term, $\mu_{4}$ has been phenomenologically determined~\cite{Deur:2014vea,Deur:2004ti, Deur:2008ej}. 
In that procedure, however, an $\alpha_s$ was assumed. Thus, using the phenomenological value would bias our extraction of the coupling toward the $\alpha_s$. 
%
Here, we will ignore $\mu_4$ and other HT since they are suppressed at the relatively high $Q^2$ covered by the EIC. 

One can obtain $\alpha_s$ from $\Gamma_1^{\rm p-n}(Q^2)$ in two ways. The first is to solve Eq.~(\ref{Eq:bjorken-SR}) for $\alpha_s$,  for each $\Gamma_1^{\rm p-n}$ data point. This maps the
$Q^2$-dependence of $\alpha_s$ but the method is inaccurate because it relies on an absolute determination of $\Gamma_1^{\rm p-n}(Q^2)$.
%
%
The second way, employed here, is more accurate and involves fitting the $Q^2$-evolution of $\Gamma_1^{\rm p-n}(Q^2)$~\cite{Altarelli:1996nm}. 
%

\section{Projection for Double-Spin Asymmetries}
We now discuss the generation process of simulated EIC data and the expected uncertainties on the double-spin asymmetries.

\subsection{Proton DIS simulation}
\label{sec:protondis}
Neutral-current $ep$ DIS events were generated using the DJANGOH 4.6.10~\cite{incl:djangoh1,incl:djangoh2} event generator, for three EIC $ep$ collision energy settings; $5{\times}41$ GeV, $10{\times}100$ GeV, and $18{\times}275$ GeV~\cite{Abdul_Khalek_2022}. 

Full details of the input parameters used in the DJANGOH generation can be found in the Appendix of Ref.~\cite{ZHANG2023168276}.
The simulated yields were scaled to provide an estimate for the total event counts that correspond to an integrated luminosity of 10~fb$^{-1}$ for each combination of energy setting and hadron polarization (longitudinal and transverse).

The operation time that corresponds to the 10~fb$^{-1}$ integrated luminosity varies with energy, which are 27.3 months, 2.7 months, and 7.8 months, respectively, for the $5{\times}41$ GeV, $10{\times}100$~GeV, and $18{\times}275$ GeV settings~\cite{Accardi:2012qut, Abdul_Khalek_2022}. In real running, 
the beam time can be split between longitudinal and transverse hadron polarization settings. Given that $A_1$ is a mainly longitudinal quantity, the relative ratio can be optimized in favor of longitudinal running.

It is also worth noting that the lowest energy EIC operations of $ep$ and $e$-$^3$He pose significant technical challenges, and solutions are being developed~\cite{Montag:2024}. Depending on the engineering and construction feasibility of the finalized accelerator design, a small adjustment is anticipated for the lowest collision energy setting. This will result in a shift in the $Q^2$ and $x$ coverage.

We chose to simulate events as detected by the
EIC Comprehensive Chromodynamics Experiment (ECCE) detector~\cite{Adkins:2022jfp}. Other proposed EIC detector designs should yield very similar results in the context discussed here. In fact, the ATHENA~\cite{ATHENA:2022hxb} and ECCE configurations have now been combined in the ePIC design. Despite differences in apparatus details, the overall kinematic range and achievable precision are expected to be similar. Generated events were passed through ECCE's GEANT4~\cite{AGOSTINELLI2003250} based full detector simulation framework to account for the impact of detector resolution, efficiency, and acceptance. The events from the output were used as pseudodata to estimate event rates for this analysis. DIS events were selected using the following criteria based on the expected performance of the ECCE detectors: 
\begin{itemize}
    \item DIS kinematics $Q^2 > 2$ GeV$^2$ and invariant mass $W>\sqrt{10}$~GeV
    \item The scattered electron's energy $E_{e}^{'} > 2$ GeV and pseudo-rapidity $\eta_{e} > -3.5$, to isolate regions of high detector efficiency
    \item The inelasticity of the scattering event $0.01 < y < 0.95$, to avoid regions of poor reconstruction/resolution and reduce the large photoproduction background in the high-$y$ region. 
\end{itemize}
Selected events were binned in two dimensions by the reconstructed values of $x$ and $Q^2$. 

One important feature of DJANGOH is that it calculates not only the vertex-level scattering for an event but also details of initial- and final-state electromagnetic and electroweak radiation for the scattering event, which would distort the event from Born-level kinematics. 
Furthermore, the energy loss of particles due to passing through material and detector resolution effects add further distortion. To correct for the sizable bin migration due to these effects, the binned events were unfolded to the Born-level distribution in $x$ and $Q^2$ with a 4-iteration Bayesian unfolding algorithm using the RooUnfold~\cite{roounfold} framework, trained with the Born-level and reconstructed values in the pseudodata. The unfolding algorithm provided an estimate of the increase of uncertainties resulting from the correction for bin migration and QED radiative effects, allowing for the estimation of both statistical and systematic uncertainties on the binned yields. 

\subsection{Neutron DIS simulation}\label{sec:neutrondis}
The neutron spin asymmetry information is traditionally extracted from the inclusive measurement of electrons scattering off light nuclei such as deuteron or $^3$He. Since the neutron spin accounts for almost 90\% of $^3$He spin, the latter is preferred for neutron spin measurement. However, the extraction introduces a sizeable systematic uncertainty due to our limited understanding of nuclear effects. The far-forward detector region at EIC provides a unique opportunity for tagging measurements in which we can detect both spectator protons from the helium-3 breakup. This double-tagging method selects the signal for scattering off a quasifree neutron in $^3$He, thereby suppressing the nuclear correction uncertainties~\cite{friscic2021neutron}.

The DJANGOH event generator was used to produce a sample of neutral-current DIS events from $^3$He, using similar input parameters to the proton-scattering case. 
As DJANGOH does not include the effects of Fermi motion, the spectator nucleons were separately generated and added to the event sample.
The distributions of the spectator nucleons were simulated using the convolution approximation for nuclear structure functions in the Bjorken limit~\cite{Frankfurt:1981mk}, using the $^3$He ground-state model of Ref.~\cite{CiofidegliAtti:2005qt}, the light-front formalism of Ref.~\cite{friscic2021neutron}, and the structure functions of Refs.~\cite{Accardi:2016qay, Accardiprivate}.

DIS $e$-$^3$He events were generated for three EIC energy settings: $5\times41$ GeV/nucleon, $10\times100$ GeV/nucleon, and $18\times166$ GeV/nucleon, and were scaled to an integrated luminosity of 10~fb$^{-1}$ for each energy and $^3$He polarization (longitudinal and transverse) setting. 
The generated events were passed through the GEANT4 simulation and analysis framework.  The first step to selecting the double spectator tagged sample is applying identical cuts on the reconstructed electron variables as applied in the $ep$ case (see Section~\ref{sec:protondis}).

In the proposed measurement, the selection criteria for spectator protons require detecting both protons remaining after the breakup of a $^3$He. This is achieved using an integrated detector stack in the far-forward region, close to the downstream hadron beamline. The goal is to favor interactions where the electron interacts with a quasifree neutron within the $^3$He. This approach, known as the "double-tagging" technique, involves detecting both spectator protons simultaneously.

After the interaction and breakup, the spectator protons continue to move along the beam momentum in the far-forward region. Since each proton acquires a different momentum difference (compared to the beam momentum) after the interaction, they are tagged by a tracking system in the far-forward region: the trackers inside the B0 dipole magnet and the Roman Pot (RP) detector. They are positioned approximately 6 m and 26 m from the interaction point (IR), respectively.

Both the B0 and RP detectors are equipped with multiple layers of capacitive couple low-gain avalanche diodes (AC-LGADs), providing an efficiency of over 95\% (per layer) and a position resolution better than 100~$\mu$m. For this study, the effective detector acceptance was considered, covering an angular range of $\pm 27$ mrad around the outgoing beam pipe. This corresponds to a pseudorapidity coverage of $4<\eta<6$.

Nuclear effects were minimized by requiring $|\vec p_{s1}+\vec p_{s2}|<0.1$ GeV, where $\vec p_{s1,2}$ are the 3-momenta of the two spectator protons, see Ref.~\cite{BYLINKIN2023168238} for details.

\begin{figure*}[!t]
    \centering
    \includegraphics[width=0.44\linewidth]{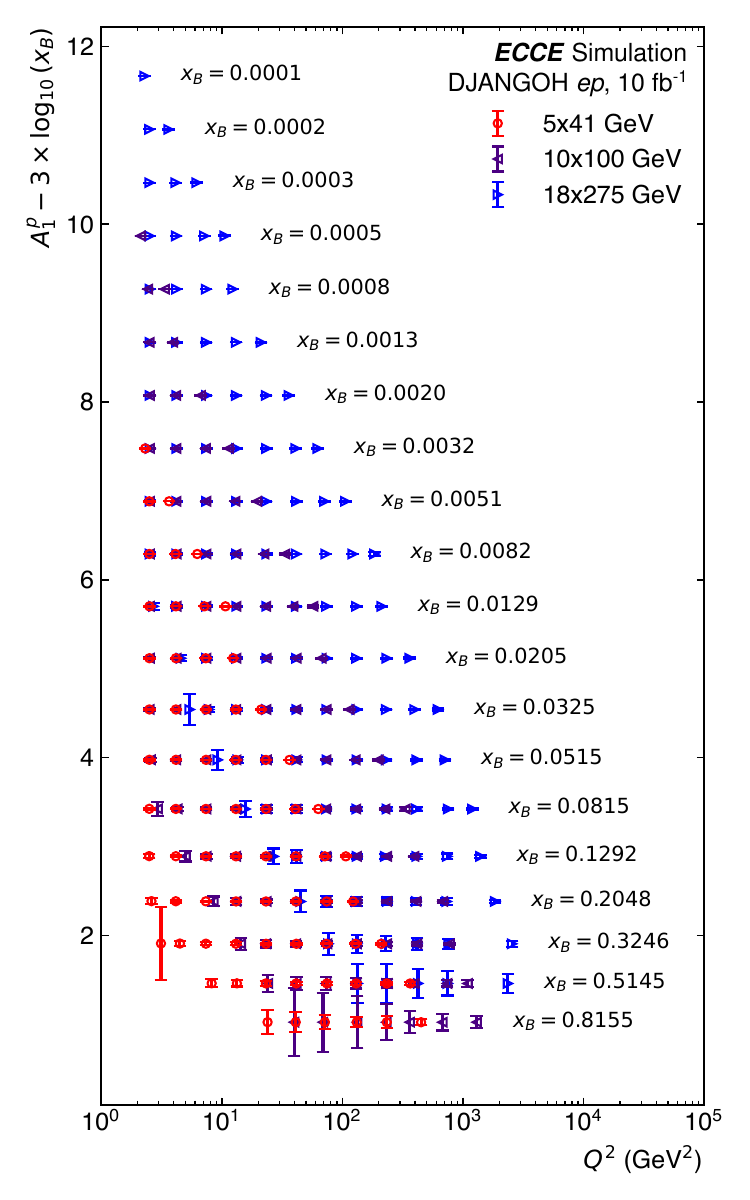}
    \includegraphics[width=0.44\linewidth]{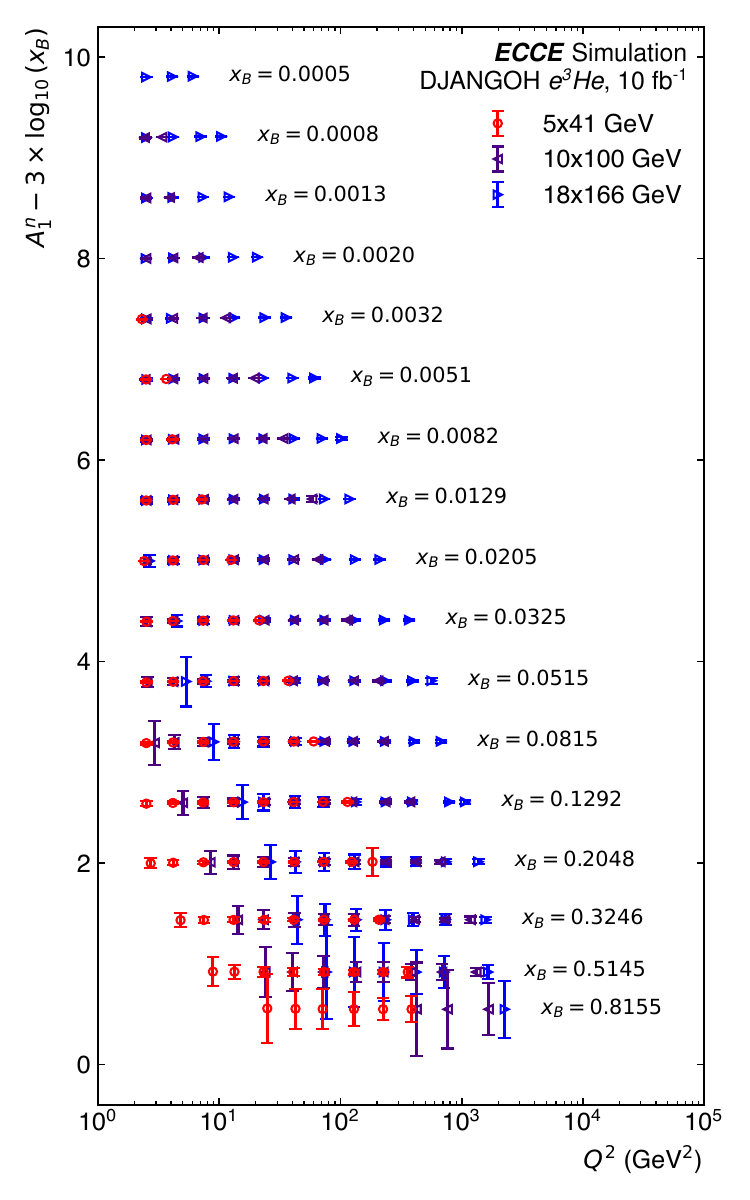}
    \caption{Coverage for $A_1^p$ and $A_1^n$ using an integrated luminosity of 10~fb$^{-1}$ for each of the longitudinally and transversely polarized hadron beam per energy settings. Three energy settings were used for each of $ep$ and $e$-$^3$He collisions. Projections for different $x$ values are by $3\times \log_{10}(x)$ for clarity. The error bars are statistical uncertainties. }
    \label{fig:A1pn}
\end{figure*}

\subsection{$A_1$ extraction}
The virtual photon asymmetry for electron scattering off a hadron of spin-1/2 is defined as
\begin{equation}
 A_1(x, Q^2)\equiv \frac{\sigma_{1/2}-\sigma_{3/2}}{\sigma_{1/2}+\sigma_{3/2}},
\end{equation}
where $\sigma_{1/2(3/2)}$ is the total virtual photoabsorption cross section for the nucleon with a projection of $1/2(3/2)$ for the total spin along the direction of the photon momentum. At high-enough $Q^2$ that satisfies $\gamma \ll 1$, where 
$
\gamma^2=(2M x)^2/Q^2
$
with $M$ the nucleon mass, $A_1\approx g_1/F_1$. Thus, we can form the spin structure function $g_1$ by combining measured $A_1$ with a world parametrization of the unpolarized structure function $F_1$.

In practice, $A_1$ is extracted from the measured longitudinal ($A_{||}$) and transverse electron asymmetries ($A_{\perp}$), 
\begin{equation*}
    A_{||}= \frac{\sigma_{\downarrow\Uparrow} 
    - \sigma_{\uparrow\Uparrow}}
    {\sigma_{\downarrow\Uparrow}
   + \sigma_{\uparrow\Uparrow}}
     \quad \text{and} \quad
    A_{\perp} = \frac{\sigma_{\downarrow\Rightarrow} - \sigma_{\uparrow\Rightarrow}}{\sigma_{\downarrow\Rightarrow} + \sigma_{\uparrow\Rightarrow}}~,
\end{equation*}
where the $\downarrow,\uparrow$ represents the spin of the longitudinally polarized electron antiparallel (parallel) to beam direction, and the $\Uparrow,\Rightarrow$ the spin of the longitudinally or transversely polarized hadron. The relation between $A_1$, $A_{||}$, and $A_{\perp}$ is 
\begin{equation}
A_{1}=\frac{A_{\|}}{D(1+\eta \xi)}-\frac{\eta A_{\perp}}{d(1+\eta \xi)},
\label{A1_formula}
\end{equation}
where~\cite{Sato:2016tuz, E99117-PRC} 
\begin{align}
    D &= \frac{y(2-y)(2+\gamma^2y)}{(2(1+\gamma^2)y^2 + (4(1-y) - \gamma^2y^2)(1+R))}\\
    d &= \sqrt{4(1-y)-\gamma^2y^2}D/(2-y)\\
    \eta &= \gamma(4(1-y) - \gamma^2y^2)/(2-y)/(2+\gamma^2y)\\
    \xi &= \gamma(2-y)/(2 + \gamma^2y)~, 
\end{align}
which are kinematic factors, except $R\equiv\sigma_{L} / \sigma_{T}$, which is the ratio of the longitudinal to transverse virtual photon absorption cross sections. The world data fit~\cite{E143:1998nvx} was used for $R$~\cite{Feynman:1973xc}.

For both the proton and $^3$He cases, the unfolded event yields in each bin of $x$ and $Q^2$ were used to estimate the yield-related uncertainties on the spin asymmetries,
\begin{equation}
\delta A_{||,\perp} \simeq \frac{\delta N}{N P_e P_N}    
\label{eqn:uncertainty}
\end{equation}
where $N$ is the total event yield in the bin, $\delta N$ is the overall uncertainty on that yield, and $P_e$ and $P_N$ are the polarizations of the electron and ion beam respectively, taken to be $(70\pm 1)$\%. 
The uncertainties $\delta A_{||,\perp}$ were then propagated into the total uncertainty on $A_1$. The kinematic coverage and uncertainties for $A_1^p$ and $A_1^n$ are shown in Fig.~\ref{fig:A1pn}.
Note that the asymmetry measurements for both parallel ($A_{||}$) and perpendicular ($A_{\perp}$) will be performed, 
removing the need to use a model for the transverse asymmetry.



\section{Extraction of $\alpha_s$  from the simulated EIC data}
\subsection{Formation of $g_1$ and its moments}\label{sec:g1}
Once the projected results on the proton and the neutron $A_1^p$ and $A_1^n$ were produced from the simulation, these were multiplied by $F_1^p$ and $F_1^n$ obtained from a global fit~\cite{Cocuzza:2022jye} to provide the projected results on $g_1^p$ and $g_1^n$. Specifically, both the central value and the uncertainty in $F_1^{p,n}$ were evaluated using the grid JAM22-STF\_proton~\cite{Cocuzza:2022jye} and its replicas. 
While we expect that future data will provide a reasonable description of the $F_1^{p,n}$ themselves, the uncertainty of the projection on the polarized quantities is dominated by the statistical uncertainty in the asymmetry projection, and thus using a world fit of the $F_1$'s should be sufficient for the present work.

The projected results on $g_1^p$ and $g_1^n$ were then integrated over the available $x$ range to obtain the integrals $\int g_1^p~dx$ and $\int g_1^n~dx$. 
For the low-$W$ (high-$x$) region, either not covered by the EIC simulated data or where the EIC projection provides large statistical uncertainty, we used a parametrization~\cite{CLAS:2017qga} that provides a good description of most of the existing world double-polarization data.  In other words, the integral of Eq.~(\ref{eq:Gamma1}) was formed as
\begin{equation}
\Gamma_1^\text{proj.} (Q^2) = \int_{x_\text{min}}^{x_\text{param.}} g_1^\text{proj.}(Q^2,x) \, dx
+  \int_{x_\text{param.}}^{1^-} g_1^\text{param.}(Q^2,x) \, dx~.\nonumber
\end{equation}
The uncertainty of the full integral accounts for the statistical uncertainties of the projection and model uncertainty, and is evaluated by varying the model inputs. The projected values of $\Gamma_1^{p,n}$ for each $Q^2$ and the coverage of EIC projection vs. model usage are summarized in Table~\ref{tab:Gamma1_proj}.

\begin{table}[!h]
    \centering
    \begin{tabular}{|c|c|c|c|c|}
        \hline
        \small{$Q^2$[GeV$^2$]}&$x_\text{min}$& $x_\text{p~param}$ & $x_\text{n~param}$ & $\Gamma_1^{p-n}$\\
        \hline
         2.37 & 0.0006 & 0.046 & 0.060 & 0.1817 $\pm$ 16 $\pm$ 06 \\
         4.22 & 0.0006 & 0.105 & 0.176 & 0.1935 $\pm$ 17 $\pm$ 10 \\
         7.50 & 0.0010 & 0.160 & 0.263 & 0.1949 $\pm$ 17 $\pm$ 13 \\
        13.34 & 0.0016 & 0.506 & 0.506 & 0.1945 $\pm$ 21 $\pm$ 23 \\
        23.71 & 0.0025 & 0.807 & 0.807 & 0.1901 $\pm$ 20 $\pm$ 23 \\
        42.17 & 0.0100 & 0.802 & 0.802 & 0.1621 $\pm$ 19 $\pm$ 21 \\
        74.99 & 0.0100 & 0.915 & 0.915 & 0.1632 $\pm$ 19 $\pm$ 19 \\
        \hline
    \end{tabular}
    \caption{Projected integrals $\Gamma_1^{p,n}$, along with the minimum $x$ covered by the EIC projection, the $x_\text{param.}$ above which the model parameterization was used for proton (third column) and neutron (fourth column), and the resulting integrals. The first uncertainty on the integral includes statistical uncertainty and experimental effects (detector smearing, bin migration, and unfolding), while the second uncertainty corresponds to that due to both electron and hadron beam polarimetry.
    }
    \label{tab:Gamma1_proj}
\end{table}

\begin{figure}[!h]
\includegraphics[width=0.45\textwidth]{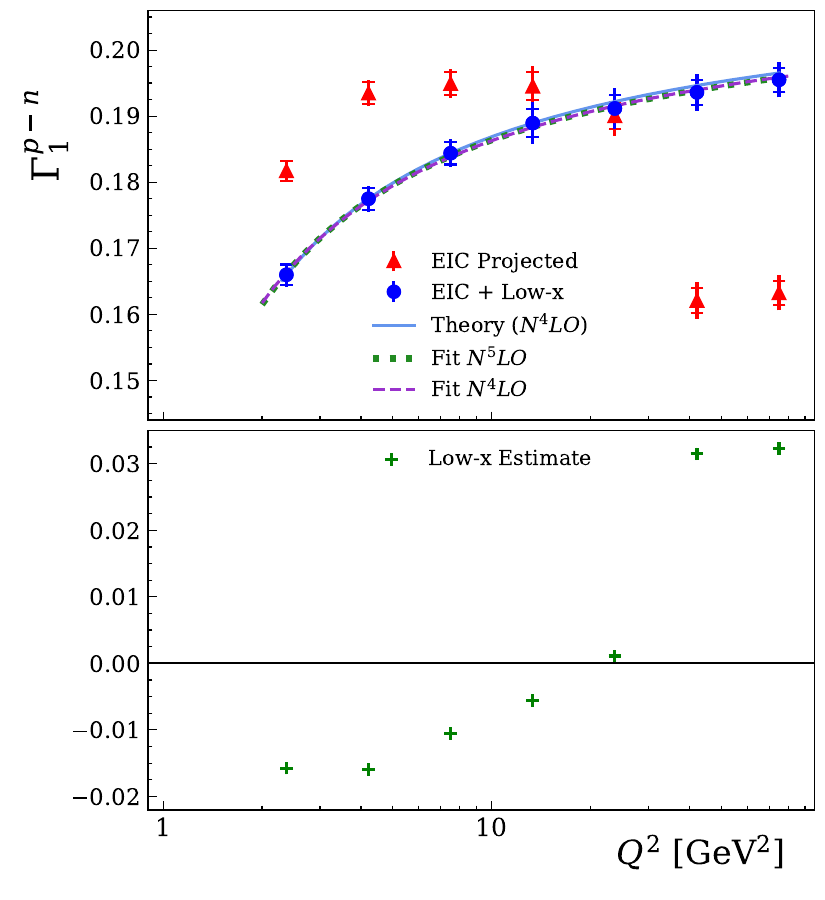}
 \caption{The Bjorken sum $ \Gamma_1^{\rm p-n}$ vs $Q^2$. The partial moment integrated over the $x$-coverage of the EIC and the high-$x$ region from the parametrization is shown as (red) solid triangles. The (blue) solid circles show the full integral, including the low-$x$ contribution shown as (green) crosses in the lower panel. 
 The inner error bars comprise the uncertainty from statistical and experimental effects (detector smearing, bin migration, and unfolding), and the outer error bar includes in addition (in quadrature) the systematic uncertainty from beam polarimetry. The (light-blue) solid curve shows the expected $\Gamma_1^{\rm p-n}(Q^2)$ calculated at N$^4$LO with $\alpha_s$ at 5-loop. 
The nearly indistinguishable (green) dotted and (purple) dashed curves are the N$^5$LO, 5-loop, and N$^4$LO, 5-loop, fits of the full integral, respectively. We do not show the uncertainty on the unmeasured low-$x$ contribution (green crosses) since it is not directly relevant to the uncertainty $\Delta \alpha_s/\alpha_s$. The relevant contribution from low-$x$ to this uncertainty is given in Table~\ref{tab:Gamma1_miss}.} 
\label{fig:bjsr-EIC}
\end{figure}

The finite beam energies and the DIS requirement of minimal $Q^2$ values limit the experimental reach at low-$x$. The unmeasured contribution,  $\Gamma_1^{\text{{low}-}x}\equiv \int_0^{x_{\text{min}}} (g_1^{\mathrm{p}} - g_1^{\mathrm{n}}) \, dx$, with $x_{\text{min}}$ being the lowest $x$ value covered by the EIC (see Table~\ref{tab:Gamma1_proj}) was estimated from the difference between the simulated partial integral and the full Bjorken integral computed with Eq.~(\ref{Eq:bjorken-SR}), see Fig.~\ref{fig:bjsr-EIC}.
This is the simplest and most accurate method to estimate the missing low-$x$ part. However, as it requires an $\alpha_s$ input for Eq.~(\ref{Eq:bjorken-SR}), it cannot be used for the actual determination of $\alpha_s$ from the future real EIC data. Yet, at the moment, the method can be used for an accurate determination of the low-$x$ uncertainty, rather than the central value of $\alpha_s$ that is immaterial for this work. For the analysis of the future EIC data, the missing low-$x$ part can be estimated using Regge theory, as was done in~\cite{Deur:2014vea}, for which parameters are expected to improve thanks to both EIC and dedicated  JLab 12 GeV experiments~\cite{Dalton:2020wdv}.
Once the value of the missing low-$x$ part was estimated, its contribution to the uncertainty $\Delta \alpha_s/\alpha_s$ was evaluated using the procedure outlined in~\cite{Deur:2014vea}:
$\Delta_Q \Gamma_1^{{\rm low}-x}=(\frac{d\Gamma_1^{\rm p-n}}{dQ^2})(\frac{\Delta Q^2}{2})(\frac{\Gamma_1^{{\rm low}-x}}{\Gamma_1^{\rm p-n}})$, with $\Delta Q^{2}$ as the $Q^2$ bin size, and the derivative $\frac{d\Gamma_1^{\rm p-n}}{dQ^2}$ calculated based on the theoretical expectation for the $\Gamma_1^{\rm p-n}$ $Q^2$-dependence. 
The unmeasured part of $\Gamma_1^{p-n}$ and $\Delta_Q \Gamma_1^{{\rm low}-x}$ are given in Table~\ref{tab:Gamma1_miss}.
\begin{table}[!h]
    \centering
    \begin{tabular}{|c|c|c|c|}
        \hline
        \small{$Q^2$[GeV$^2$]}& $x_\text{min}$ & $\int_{~0}^{x \rm min} g_1^{p-n} dx$ & $\Delta_Q \Gamma_1^{{\rm low}-x}$ \\
        \hline
         2.37 & 0.0006 & $-15.74\times10^{-3}$  & $0.60\times10^{-3}$  \\
         4.22 & 0.0006 & $-15.98\times10^{-3}$  &  $0.30\times10^{-3}$  \\
         7.50 & 0.0010 & $-10.54\times10^{-3}$  & $0.12\times10^{-3}$  \\
        13.34 & 0.0016 & $-5.57\times10^{-3}$  & $0.04\times10^{-3}$  \\
        23.71 & 0.0025 & $1.11\times10^{-3}$  & $0.00\times10^{-3}$  \\
        42.17 & 0.0100 & $31.50\times10^{-3}$  & $0.13\times10^{-3}$  \\
        74.99 & 0.0100 & $32.26\times10^{-3}$  & $0.11\times10^{-3}$  \\
        \hline
    \end{tabular}
    \caption{ Unmeasured low-$x$ part of $\Gamma_1^{p-n}$.
    The fourth column provides the uncertainty on the normalized $Q^2$-dependence of the missing part (third column), see main text.
    }
    \label{tab:Gamma1_miss}
\end{table}

\subsection{Extraction of $\alpha_S$}
To extract $\alpha_s(M_{Z^0}^2)$, the simulated $\Gamma_1^{\rm p-n}(Q^2)$ values are fitted using  Eq.~(\ref{Eq:bjorken-SR}).
The key free parameter in the fit is the QCD scale  
$\Lambda_s$~\cite{Deur:2023dzc}. Knowing it
is sufficient to provide the value of $\alpha_s(M_{Z^0}^2)$. 
For instance, at LO, $\alpha_s(Q^2)=\frac{12\pi}{(11n_c-2n_f) \ln\big(Q^2/ \Lambda_s^2\big)}$, with $n_c$ and $n_f$ the numbers of quark colors and flavors, respectively.
From the fit, $\Delta \Lambda_s$ is derived, which allows the computation of $\Delta \alpha_s(M_{Z^0}^2)$, the quantity of interest in this study.

The other free parameter in the fit is $g_A$. Despite being well-measured, we allow it to vary freely to accommodate potential offsets from the systematic uncertainties in the $\Gamma_1^{\rm p-n}$ 
expected in the real 
data. 

Generally, the total uncertainty is minimized by optimizing the number of low-$Q^2$ points versus high-$Q^2$ points. The former provides higher sensitivity to $\alpha_s$ and minimal fit uncertainty but increases truncation uncertainty. The latter becomes asymptotically insensitive to $\alpha_s$ and exhibits increased uncertainties at low-$x$. 
For this work, we found that including all simulated data, viz. fitting over a range of $2.4<Q^2<75$ GeV$^2$, is optimal. 

Using N$^4$LO and LT, our best-fit yields $\Delta \alpha_s/\alpha_s = [\pm 8.3(\text{fit}) \pm 6.4(\text{trnc})]\times10^{-3}$, where the first uncertainty arises from the fit, and the second from the truncation of the pQCD series, Eq.~(\ref{Eq:bjorken-SR}). The uncertainty of the fit accounts for the expected statistical uncertainty of the measurement, experimental effects (detector smearing, bin migration, and unfolding), the missing low-$x$ contribution, and the parametrizations.
%
The truncation uncertainty is estimated by refitting the data with the $\Gamma_1^{\text{p-n}}$ and $\alpha_s$ approximants at N$^5$LO and taking the difference $|\Lambda_s^{(5)} - \Lambda_s^{(4)}|/2$ as the truncation error. [Here the superscript $(4)$ or $(5)$ denotes the order for the series of $\Gamma_1^{\rm p-n}$ and $\alpha_s$.]

To account for the beam polarization uncertainty, we utilized a Monte Carlo-based method. First, we generated pseudodata for $g_1^{p,n}$ of each beam energy settings $i$, with the centroid given by the parametrizations as described in Sec.~\ref{sec:g1}:
\begin{eqnarray}
  g_{1,i}^{\rm{MC,syst}} = g_1^{\rm{centroid}}(1+r_i)~,
\end{eqnarray}
where $r_i$ is a random number following a normal distribution of standard deviation 0.02 to account for the 2\% combined relative uncertainty of the two beam polarizations, with $i=1,2,\dots 6$ standing for the two beam types ($p$ or $^3$He) and the three energy settings. The $r_i$ is independent (uncorrelated) among the six settings. The generated $g_1$ are then combined to form $\Gamma_1^{p-n}$ following the procedure of Section~\ref{sec:g1}. 
In the region where data from two beam energies overlap, their statistical uncertainties are combined, and the systematic uncertainties are combined following the same statistical weighting.  The resulting uncertainty on $\Gamma_1^{p-n}$ is shown in Table~\ref{tab:Gamma1_proj}. 
The $\Gamma_1^{p-n}$ from each Monte Carlo event is then fed to the $\alpha_s$ fitting procedure with all other uncertainties (statistical, truncation, low-$x$) set to zero, 
such that fluctuations in the fitted $\alpha_s$ result would represent the effect from the beam polarization uncertainty.

This yields an uncertainty on $\alpha_s$ of 0.78\%. Another source of systematic uncertainty is radiative corrections, which were taken into account in the unfolding procedure described in Sec.~\ref{sec:protondis}.

\begin{figure}[!h]
\includegraphics[width=0.45\textwidth]{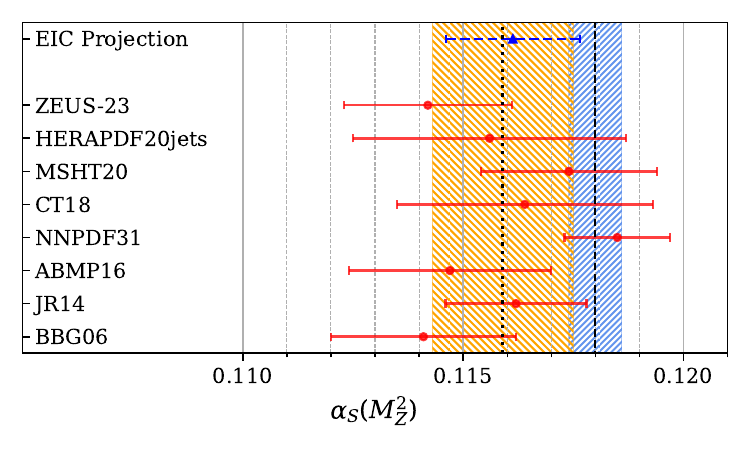}
\caption{Expected total uncertainty on $\alpha_s$ from this study (blue triangle with dashed error bar) compared with existing results from DIS world data (red solid circles with solid error bars)~\cite{ParticleDataGroup:2020ssz, ZEUS:2023zie}. The Particle Data Group~\cite{ParticleDataGroup:2020ssz} average of DIS PDF fits is shown as the (golden) negative-slope hatched band on the left, while the world average combining all data and lattice QCD results is shown as the (blue) positive-slope hatched band on the right.}\label{fig:alpha_s}
\end{figure}

Adding all uncertainties in quadrature, 
we obtain a precision on $\alpha_s$ of {$\Delta \alpha_s/\alpha_s = 1.3\%$}. 
The simulated $\Gamma_1^{\text{p-n}}(Q^2)$ and best fits are shown in Fig.~\ref{fig:bjsr-EIC}. 
Also shown is the missing low-$x$ contribution (lower panel). Notably, it becomes negative for $Q^2 \lesssim 23$~GeV. There is no {\it a priori} reason to expect the low-$x$ missing contribution to be positive since the BJSR components need not be positive quantities and since the value of the missing low-$x$ part depends on experimental conditions. Yet, it is likely that the negative low-$x$ contribution at the lower $Q^2$ stems from uncertainties in the PDFs used to estimate the $\Gamma_1$. In fact, the missing contribution is small since $x_{min}$$\sim$$10^{-3}$. A Regge theory estimate~\cite{Bass:1997fh} predicts it to be about 2\% of the expected full $\Gamma_1^{p-n}$. Thus, PDF systematics may affect the sign of the missing contribution if the partial $\Gamma_1^{p-n}$ integral is overestimated by a few \%. At larger $Q^2$, the missing low-$x$ part is larger, about  20\% using~\cite{Bass:1997fh}, so its sign is not affected by PDF uncertainties. 
The expected uncertainty on $\alpha_s$, shown in Fig.~\ref{fig:alpha_s}, provides a competitive level of precision compared with the current 1.7\% uncertainty derived from DIS world data, primarily obtained through global PDF fits of unpolarized and polarized structure functions~\cite{ParticleDataGroup:2020ssz}. 

Another important 
uncertainty is the uncertainty in determining the relative luminosity difference between two spin states of the longitudinally polarized electron beam (parallel and antiparallel to beam direction). According to Ref.~\cite{Abdul_Khalek_2022}, the EIC physics program requires this uncertainty to be below $10^{-5}$ to benefit from the statistics of EIC, given the spin asymmetry at low-$x$ is already at the $10^{-4}$ level.  
In our study, we in fact found the effect on the Bjorken sum to be negligible when a $10^{-4}$ luminosity uncertainty is included, compared with the uncertainty shown in Table~\ref{tab:Gamma1_proj}. 
In addition, the uncertainty from the luminosity difference can be mitigated by reversing the beam polarization at the source level, a common practice at, e.g., JLab, which we expect will be available at the EIC.

\subsection{Discussion}

The truncation uncertainty being important, we explored using the principle of maximum conformality (PMC) method~\cite{Wang:2023ttk} to reduce that uncertainty. The PMC optimizes a perturbative approximant, thereby allowing for a more accurate determination of $\alpha_s$ from the corresponding observable.
The PMC extends the well-known Brodsky-Lepage-Mackenzie method~\cite{Brodsky:1982gc}, which optimizes perturbative predictions by summing all $\beta$-terms into $\alpha_s(Q^2)$. The PMC series for $\Gamma_1^{\rm p-n}(Q^2)$ is calculated in~\cite{Deur:2017cvd}. In this series, each N$^n$LO order is characterized by a specific scale, $Q^2_n$, that matches the virtuality in the order-$n$ processes.
Unfortunately, in the case of $\Gamma_1^{\rm p-n}(Q^2)$, the NLO PMC scale is very small, $Q^2_2 \simeq 0.047Q^2$~\cite{Deur:2017cvd}, which precludes using data below $Q^2 \simeq 20$ GeV$^2$. We find that excluding these data outweighs the improvement in truncation uncertainty.

The BJSR is just one method to determine $\alpha_s$. Other methods, such as global PDF fits, are also viable at the EIC and should also significantly enhance our determination of $\alpha_s$ using DIS data.
Additionally, the optimal range of the fit requires the lowest-$Q^2$ point, which means that EIC data can be fruitfully complemented by lower $Q^2$ data with reasonably low-$x$ coverage, such as from the possible energy upgrade of JLab. The combination of EIC and JLab data at 22 GeV could potentially yield $\Delta \alpha_s/\alpha_s=0.6\%$~\cite{Accardi:2023chb}. 

It is worth noting that at the EIC, the neutron spin structure could be measured using a polarized deuteron beam. The extraction of the neutron spin structure from inclusive DIS via $e$D with the single spectator proton tagged) will have a higher statistical precision than for double-tagged $e-^3$He given the same luminosity~\cite{Frankfurt:1981mk}.

It is understood that preserving the polarization of the polarized deuteron in a circular storage ring is challenging and is beyond the scope of the planned spin rotation capabilities of the EIC. Currently, the development of a polarized deuteron beam is not included in the EIC project research and development effort. It is expected that the experience gained from the polarized $^3$He beam (projected to be optional in the initial years of EIC), will directly benefit the development of the deuteron beam and polarimetry~\cite{Abdul_Khalek_2022}. The polarized deuteron beam will be available in the form of a future upgrade to the EIC project.

\section{Conclusion}
The BJSR offers several advantages for an experimental extraction of $\alpha_s$. It has the robustness of inclusive data, the simplest pQCD evolution (the $x$-dependence is integrated out and the isospin nature of the BJSR fully or partially suppresses contributions from difficult quantities such as $\Delta G$ or coherent reactions), and crucially is model independent; the sum rule encapsulates the nonperturbative part of the reaction into the well-measured axial charge $g_A$. The missing low-$x$ issue, which is often a significant challenge in studies involving moments, is by design minimal in the case of the EIC. 
A realistic simulation of the EIC data demonstrates that the BJSR can yield $\alpha_s(M_{Z^0}^2)$ with an accuracy of approximately $\Delta \alpha_s/\alpha_s \simeq 1.3\%$. 
This result compares well with the best current extractions of $\alpha_s$ from experimental data. Furthermore, independent methods to accurately determine $\alpha_s$ will be possible using other types of EIC data, e.g., inclusive neutral and charged current reactions studied in~\cite{Cerci:2023uhu}. 
 
These approaches will achieve a $\Delta \alpha_s/\alpha_s$ below the percentage level, which would make the EIC a key contributor to the global extraction of $\alpha_s$.

\noindent \textbf{ACKNOWLEDGMENTS} This material is based upon work supported by the U.S. Department of Energy (DOE), Office of Science, Office of Nuclear Physics under Contract No. DE-AC05-06OR23177. W.B.~Li and A.~Deshpande were supported by the U.S. DOE Award No. DE-1020FG02-05ER41372 and the Center for Frontiers in Nuclear Science (CFNS), Stony Brook University. The work of D.W. Upton, C. Cotton, M. Nycz, and X. Zheng was supported by the U.S. DOE Award No. DE–SC0014434. Jefferson Science Associates LLC operates the Thomas Jefferson National Accelerator Facility for the DOE under Contract No. DE-AC05-84ER40150, mod. \# 175. The work of J.R. Pybus and M. Nycz was supported in part by the EIC$^2$ Center fellowship at Jefferson Lab. We thank Dave Gaskell for useful discussions regarding beam polarimetry measurements. We also extend our gratitude to Katarzyna Wichmann and Barak Schmookler for their comments, which have greatly enhanced the quality of this paper.

\bibliography{a_EIC.bib}

\end{document}

%% file: review_tool.tex
\usepackage{mfirstuc} 
\newcommand{\addReviewer}[2]{
  \expandafter\newcommand\csname #1\endcsname[1]{{\textbf{ \color{#2} \capitalisewords{#1}:\,##1}}}
  \expandafter\newcommand\csname #1cor\endcsname[2]{{\color{#2} \capitalisewords{#1}:\,\st{##1}{\textbf{##2}}}}
  \expandafter\newcommand\csname #1color\endcsname{#2}
  \expandafter\newcommand\csname #1todo\endcsname[1]{{\todo[inline,color=white!70!#2, caption={}]{\textbf{\capitalisewords{#1}}: ##1}}}
}

\usepackage{tikz, marginnote} 
\newcommand{\checkedby}[1]{
\ifdefined\CROSSCHECKS
  \marginnote{
    \begin{tikzpicture}
      \foreach \x [count=\xi] in {#1} {
         \node[shape=circle,inner sep=0mm,
         minimum size=2mm,
         fill=\csname \x color\endcsname] at (\xi*3mm,0) {};
       }
    \end{tikzpicture}
  }
\else
\fi
}

\usepackage{soul,color}
\definecolor{chromeyellow}{rgb}{1.0, 0.65, 0.0}
\definecolor{DodgeBlue}{rgb}{0.118, 0.565,1.000}
\definecolor{asparagus}{rgb}{0.53, 0.66, 0.42}
\definecolor{cadmiumgreen}{rgb}{0.0, 0.42, 0.24}
\definecolor{blue(ryb)}{rgb}{0.01, 0.28, 1.0}
\definecolor{periwinkle}{RGB}{181, 146, 203}
\definecolor{turquoiseblue}{rgb}{0.02, 0.55, 0.55}
\definecolor{atlanticstorm}{RGB}{57,107,127}
\definecolor{amethyst}{rgb}{0.6, 0.4, 0.8}
\definecolor{blue-violet}{rgb}{0.54, 0.17, 0.89}
\definecolor{brightlavender}{rgb}{0.75, 0.58, 0.89}
\definecolor{brightmaroon}{rgb}{0.66, 0.13, 0.28}
\definecolor{brilliantrose}{rgb}{1.0, 0.33, 0.64}
\definecolor{byzantine}{rgb}{0.74, 0.2, 0.64}
\definecolor{darkmagenta}{rgb}{0.55, 0.0, 0.55}

\addReviewer{alessandro}{brightmaroon}
\addReviewer{someone}{asparagus}
\addReviewer{AD}{red}
\addReviewer{xiaochao}{blue}
\addReviewer{TK}{purple}

\addReviewer{bob}{chromeyellow}
\addReviewer{vincent}{brown}
\addReviewer{patrizia}{magenta}
\addReviewer{andrew}{purple}
\addReviewer{arkaitz}{asparagus}
\addReviewer{daniel}{periwinkle}
\addReviewer{eugene}{byzantine}
\addReviewer{natalia}{brightmaroon}



%% file: authors_alphaS_rev.tex
\author{T.~Kutz}
\affiliation{Massachusetts Institute of Technology, Cambridge, Massachusetts, USA}

\author{J.~R.~Pybus}
\affiliation{Massachusetts Institute of Technology, Cambridge, Massachusetts, USA}

\author{D.~W.~Upton}
\affiliation{University of Virginia, Charlottesville, Virginia, USA}
\affiliation{Old Dominion University, Norfolk, Virginia, USA}

\author{C.~Cotton}
\affiliation{University of Virginia, Charlottesville, Virginia, USA}

\author{A.~Deshpande}
\affiliation{Center for Frontiers in Nuclear Science, Stony Brook, New York, USA}
\affiliation{Stony Brook University, Stony Brook, New York, USA}

\author{A.~Deur}
\email{deurpam@jlab.org}
\affiliation{Thomas Jefferson National Accelerator Facility, Newport News, Virginia, USA}
  
\author{W.B.~Li}
\affiliation{Center for Frontiers in Nuclear Science, Stony Brook, New York, USA}
\affiliation{Stony Brook University, Stony Brook, New York, USA}
\affiliation{Mississippi State University, Starkville, Mississippi, USA}
\affiliation{The College of William and Mary, Williamsburg, Virginia, USA} 

\author{D.~Nguyen}
\affiliation{Thomas Jefferson National Accelerator Facility, Newport News, Virginia, USA}
\affiliation{University of Tennessee, Knoxville, Tennessee, USA}
  
\author{M.~Nycz}
\affiliation{University of Virginia, Charlottesville, Virginia, USA}
 
\author{X.~Zheng}
\affiliation{University of Virginia, Charlottesville, Virginia, USA}